\documentclass[showpacs,showkeys,twocolumn,preprintnumbers,nofootinbib,groupedaddress,superscriptaddress,amsmath,amssymb,showabstract]{revtex4-1}

\usepackage{graphicx}
\usepackage{dcolumn} 
\usepackage{bm}
\usepackage{amssymb}
\usepackage{amsmath}
\usepackage{epsfig}    
\usepackage[dvipsnames]{xcolor}
\usepackage{slashed}
\usepackage{hyperref} 
\usepackage{multirow}
\begin{document} 

\title{\boldmath \color{BrickRed} Probing Spontaneous CP-Violation through Precision Higgs Observables}

\author{Tanmoy Mondal}
\email{tanmoy.mondal@pilani.bits-pilani.ac.in}
\affiliation{Department of Physics, Birla Institute of Technology and Science, Pilani, 333031, Rajasthan, India}

\author{Kei Yagyu}
\email{yagyu@rs.tus.ac.jp}
\affiliation{Department of Physics, Tokyo University of Science,
1-3, Kagurazaka, Shinjuku-ku, Tokyo 162-8601, Japan}
%
%

\begin{abstract}
\noindent
We investigate the implications of spontaneous CP-violation in 
the general two Higgs doublet model, which leads to a non-decoupling structure of the Higgs sector. All the masses of the Higgs bosons are purely determined by the vacuum expectation 
value of the Higgs fields, and are thus constrained to be smaller than ${\cal O}(500)$ GeV by the perturbative unitarity bound.
 Such a non-decoupling nature predicts sizable deviations from the standard model expectations in the observables of the discovered Higgs boson ($h$). We find that the magnitude of deviations in ${\cal B}_{h \to \gamma\gamma}$ (${\cal B}_{h \to Z\gamma}$) are larger than 
   $\sim 10\%~(4\%)$ in the Higgs alignment limit. Moreover, we show that a robust correlation emerges between the deviations in the one-loop corrected $hhh$ coupling and ${\cal B}_{h \to \gamma\gamma}$ to be, e.g., 200\%~(50\%) and $-10.3\%$ ($-10.8\%$), respectively, under the constraints from theoretical bounds and current experimental data.  
   Using a few benchmark points, we highlight that flavor-violating decays of the additional Higgs 
   bosons can be sizable due to the constrained structure of the Yukawa interactions.

\end{abstract}
\maketitle 

\noindent
{\it Introduction} -- CP-violation (CPV) observed in decays of $K$ and $B$ mesons can successfully be explained by the three-generation structure in the Standard Model (SM) of particle physics~\cite{Charles:2004jd}. 
In the SM, the CP-violating phase or equivalently the Kobayashi-Maskawa (KM) phase is originated from complex phases of the Yukawa couplings of quarks~\cite{Kobayashi:1973fv}. Such CPV is so-called the explicit violation as the Lagrangian breaks the CP-invariance. 

There is an alternative possibility to explain the CPV, i.e., spontaneous CP-violation (SCPV), where the Lagrangian respects the CP-invariance, but the vacuum does not. SCPV cannot be realized in the SM, because the Vacuum Expectation Value (VEV) is taken to be real without loss of generality. 
However, if the Higgs sector is extended from the minimal form with one isospin scalar doublet, the SCPV is possible via complex phases of VEVs. 
Since the true structure of the Higgs sector is still unknown, the origin of CPV, i.e.,  explicit or spontaneous, also remains an open question. Therefore, determination of the nature of CPV is quite important for uncovering the structure of the Higgs sector, and to clarify the baryon asymmetry of the Universe because the CPV is one of the necessary ingredients~\cite{Sakharov:1967dj}. 

The SCPV has originally been proposed by T.~D.~Lee~\cite{Lee:1973iz} in 2 Higgs doublet models (2HDMs). 
The necessary and sufficient condition for the SCPV has been clarified in Ref.~\cite{Gunion:2005ja} in a basis independent way, i.e., at least one of three $J$-invariants~\cite{Lavoura:1994fv,Grzadkowski:2014ada} is non-zero and all the four $I$-invariants are zero. 
See also~\cite{Grzadkowski:2016szj} for the physical conditions of SCPV. 
In 2HDMs, flavor alignments are often  imposed to avoid flavor-changing neutral currents (FCNCs) via Higgs boson mediations
at tree level, which can be realized by introducing a $Z_2$ symmetry~\cite{Glashow:1976nt} or Yukawa alignment~\cite{Pich:2009sp}. However, such a scenario cannot generate the KM phase, because the CP-phase from the Higgs VEV can be removed by rephasing quark fields~\cite{Nierste:2019fbx}. 
We thus consider the 2HDM without imposing any additional symmetry, except for the CP-invariance of the Lagrangian.

One of the striking features of the SCPV 2HDM is seen in masses of Higgs bosons which are purely given by the VEV, i.e., all the dimensionful parameters in the potential are eliminated after solving the vacuum conditions. Thus, there are upper limits on the masses provided by perturbative unitarity or perturbativety~\cite{Barenboim:2001vu,Nebot:2018nqn,Nebot:2019qvr,Nierste:2019fbx,Miro:2024zka}. 
Such structure of the Higgs masses leads to non-decoupling nature, in which
we obtain sizable and correlated deviations in observables of the discovered Higgs boson ($h$) from the corresponding SM predictions~\cite{Bhattacharyya:2014oka,Arco:2023sac}. In particular, the trilinear Higgs coupling, $\lambda_{hhh}$, receives large quantum corrections induced by extra Higgs bosons, while the decay rates of $h \to \gamma\gamma$ and $h \to Z\gamma$ are suppressed due to charged Higgs boson loops, yielding a robust negative correlation between these quantities. 
We show that there is a one-to-one correspondence between ${\cal B}_{h \to \gamma\gamma}$ and the charged Higgs mass $(m_{H^\pm})$ in the Higgs alignment limit, and it gives the lower limit on $m_{H^\pm}$ to be around 220 GeV from the current Higgs signal strength measured at LHC. 
On the other hand, the upper limit on $m_{H^\pm}$ is given to be around 510 GeV coming from the combined constraints by theoretical bounds and current experimental data. 
We find that the measurement of $\lambda_{hhh}$ at the High-Luminosity LHC (HL-LHC) will put a stronger upper limit on $m_{H^\pm}$ compared to any theoretical bounds.

In addition to the physics of $h$,  the constrained structure of the Yukawa couplings are predicted in this scenario, and gives rise to flavor-violating decays of the extra Higgs bosons, offering complementary discovery channels.
Therefore, the SCPV could be explored at collider experiments such as the HL-LHC or future lepton colliders.

\noindent
{\it Model} -- We parameterize isospin scalar doublets $\Phi_{1,2}$ as
\begin{align}
\Phi_a= e^{i\xi_a}
\begin{pmatrix}
\omega_a^+\\
\frac{v_a+h_a+iz_a}{\sqrt{2}}
\end{pmatrix},\hspace{3mm}(a=1,2), \label{Eq:parametrizations}
\end{align}
where $v_a \in \mathbb{R}$ are the VEVs and the phases $\xi_a$ are taken to be $(\xi_1,\xi_2)=(0,\xi)$ using the global hypercharge symmetry without loss of generality.
As we define our Lagrangian in a basis where all the parameters are taken to be real,  
the phase $\xi$ ($\neq n\pi,~n\in \mathbb{Z}$) breaks the CP symmetry spontaneously. 
In Refs.~\cite{Ferreira:2004yd,Barroso:2005sm,Barroso:2007rr}, it has been proven that when the spontaneously CP-breaking minimum exists, it is unique, and there is no tunneling to the other types of minima such as an electric charge-breaking one or a normal one, having two real VEVs from neutral components of two Higgs doublets.

It is convenient to define the so-called Higgs basis~\cite{Davidson:2005cw} as
\begin{align}
\left(\begin{array}{c}
\Phi_1\\
\Phi_2
\end{array}\right)=
\begin{pmatrix}
1 & 0\\
0 & e^{i\xi}
\end{pmatrix}
\left(\begin{array}{cc}
c_\beta & -s_\beta\\
s_\beta & c_\beta
\end{array}\right)
\left(\begin{array}{c}
\Phi\\
\Phi'
\end{array}\right), \label{eq:hb2}
\end{align}
where $t_\beta = v_2/v_1$, 
$t_X^{} =\tan X$, $c_X^{} = \cos X$, $s_X^{} = \sin X$, and 
\begin{align}
\Phi=
\begin{pmatrix}
G^+\\
\frac{h_1'+v+ iG^0}{\sqrt{2}}
\end{pmatrix},~~
\Phi'=
\begin{pmatrix}
H^+\\
\frac{h_2'+ih_3'}{\sqrt{2}}
\end{pmatrix}, \label{Higgs-basis}
\end{align}
with $v=\sqrt{v_1^2+v_2^2}\simeq 246$ GeV.
In the above expression, $G^\pm$ and $G^0$ are the Nambu-Goldstone bosons which are absorbed into the longitudinal component of $W^\pm$ and $Z$, respectively, while 
$H^\pm$ $(h_{1,2,3}')$ are the physical singly-charged (neutral) Higgs bosons.  
The neutral Higgs bosons can be mixed with each other depending on the parameters in the Higgs potential.

The Higgs potential with explicit CP-conservation can generally be written as 
\begin{align}
&V=m_1^2|\Phi_1|^2 + m_2^2|\Phi_2|^2-(m_3^2\Phi_1^\dagger \Phi_2 +\text{h.c.})\notag\\
&+\frac{\lambda_1}{2}|\Phi_1|^4+\frac{\lambda_2}{2}|\Phi_2|^4+\lambda_3|\Phi_1|^2|\Phi_2|^2+\lambda_4|\Phi_1^\dagger\Phi_2|^2\notag\\
& 
+ \left[\frac{\lambda_5}{2}(\Phi_1^\dagger\Phi_2)+\lambda_6|\Phi_1|^2+\lambda_7|\Phi_2|^2\right](\Phi_1^\dagger\Phi_2)+\text{h.c.}, \label{pot_thdm1}
\end{align}
where all the parameters are real. 
We can write down the potential in the Higgs basis by replacing 
$(\Phi_1,\Phi_2,m_i^2,\lambda_j)\to (\Phi,\Phi',M_i^2,\Lambda_j)$ with $i=1$--3 and $j=1$--7 in Eq.~(\ref{pot_thdm1}), where the new parameters $M_i^2$ and $\Lambda_j$ are given as a function of $m_{1\text{-}3}^2$ and $\lambda_{1\text{-}7}$, respectively. Differently from the case with  explicit CPV, $M_i^2$ and $\Lambda_j$ are not completely independent due to the reality of the original parameters defined in Eq.~(\ref{pot_thdm1}).
We also note that some of the potential parameters become complex in the Higgs basis, but it does not correspond to the explicit CPV because a basis still exists in which all Lagrangian parameters are real~\cite{Gunion:2005ja}.
 
By solving the following vacuum conditions, 
\begin{align}
m_1^2 &= -\frac{v^2}{2}  (\lambda_1c_\beta^2 +\lambda_3s_\beta^2 -\lambda_4s_\beta^2 -\lambda_5s_\beta^2 + \lambda_6 s_{2\beta} c_\xi),\\
m_2^2 &= -\frac{v^2}{2} (\lambda_2s_\beta^2 +\lambda_3c_\beta^2 +\lambda_4c_\beta^2 -\lambda_5c_\beta^2 + \lambda_7 s_{2\beta}  c_\xi),\\
m_3^2&=\frac{v^2}{2}  (\lambda_5 s_{2\beta} c_\xi +  \lambda_6c_\beta^2 + \lambda_7s_\beta^2), 
\end{align}
all the dimensionful parameters, $m_i^2$, are constrained to be the order of ${\cal O}(v^2)$, 
so that all the Higgs boson masses are given by the Higgs VEV as seen below. 
The mass of charged Higgs boson is given by 
\begin{align}
m_{H^\pm}^2 & = \frac{v^2}{2}(\lambda_5 - \lambda_4). \label{eq:mch}
\end{align}
The mass matrix for the neutral Higgs bosons ${\cal M}$ can be written by the symmetric $3\times 3$ form whose elements are given in the Higgs basis, i.e., ($h_1',h_2',h_3'$) in Appendix~\ref{app:a}. 
The mass eigenstates are defined as 
\begin{align}
\begin{pmatrix}
h_1' \\
h_2' \\
h_3' 
\end{pmatrix} 
= 
R
\begin{pmatrix}
H_1 \\
H_2 \\
H_3 
\end{pmatrix}, \label{eq:rot}
\end{align}
where $R$ is the $3\times 3$ orthogonal matrix determined as 
\begin{align}
R^T {\cal M} R = \text{diag}(m_{H_1}^2,m_{H_2}^2,m_{H_3}^2).  
\end{align}
We identify the $H_1 (= h)$ state as the discovered Higgs boson with the mass of 125 GeV. 
The additional Higgs bosons $H_2$ and $H_3$ can be lighter than $h$ in general, but we consider the case where the masses of $H_{2,3}$ as well as $H^\pm$ are larger than 125 GeV throughout the Letter.

Taking the determinant of $\cal M$, we obtain 
\begin{align}
\text{det}[{\cal M}]\propto\frac{v^6 s_\xi^2}{(t_\beta + 1/t_\beta)^2}, 
\end{align}
where the proportional factor is given by a function of $\lambda$'s. 
Since values of $\lambda_i$ are constrained by perturbative unitarity bounds and/or perturbativity conditions, the above expression tells us the existence of upper limits on the masses of additional Higgs bosons, i.e., 
the decoupling limit does not exist.  
Also, 
the product of the three masses becomes zero if we take $s_\xi \to 0$, $t_\beta \to 0$ or $t_\beta \to \infty$, which provides
the lower limit on $|s_\xi|$ and the lower and upper limit on $|t_\beta|$. See e.g., Ref.~\cite{Nierste:2019fbx} for bounds using the NLO perturbative unitarity.

Generally, $h$ can be mixed with the other two neutral states $H_{2,3}$, by which the $h$ couplings are modified from the SM predictions at tree level. Since the current LHC data show that the properties of the discovered Higgs boson are consistent with the SM ones~\cite{ATLAS:2022vkf}, scenarios with an approximate Higgs alignment are favored. 
In the Higgs alignment limit, the $R$ matrix is taken to be 
\begin{align}
R \xrightarrow[{\rm Higgs~alignment}]{}
\begin{pmatrix}
1 & 0 & 0 \\
0 & c_\alpha & -s_\alpha \\
0 & s_\alpha & c_\alpha
\end{pmatrix}. 
\end{align}
This is realized by taking 
\begin{align}
{\cal M}_{12} =  {\cal M}_{13} = 0,  
\end{align}
and they determine $\lambda_{6,7}$ in terms of the other quartic couplings, see (\ref{eq:ha}).


Next, we discuss the Yukawa interactions. Generally, we have two independent $3\times 3$ Yukawa matrices $Y_1^f$ and $Y_2^f$ ($Y_{1,2}^f \in \mathbb{R}$) for $\Phi_1$ and $\Phi_2$, respectively, for each fermion type $f$ with $f$ being up-type quarks, down-type quarks and charged leptons. 
Without loss of generality, the Yukawa interactions are written in the Higgs basis and the mass eigenbasis of fermions as 
\begin{align}
\mathcal{L}_Y &=
-\Bigg[\overline{Q_L^d}\left(\frac{\sqrt{2}M_d}{v} \Phi + \rho_d\Phi'\right)d_R^{}\notag\\
&+\overline{Q_L^u}\left(\frac{\sqrt{2}M_u}{v} \Phi^c + \rho_u \Phi^{\prime c}\right)u_R^{} \notag\\
&+\overline{L_L}\left(\frac{\sqrt{2}M_e}{v}\Phi+\rho_e\Phi'\right) e_R^{}
+\text{h.c.}\Bigg],  \label{BottomUp:2HDM:Yukawa2}
\end{align}
where $M_f$ ($f=u,d,e$) are the diagonalized mass matrices with real-positive eigenvalues, and
\begin{align}
Q_L^d = 
\begin{pmatrix}
V^\dagger u_L \\
d_L
\end{pmatrix},\quad
Q_L^u = 
\begin{pmatrix}
u_L \\
V d_L
\end{pmatrix}, 
\end{align}
with $V$ being the Cabibbo-Kobayashi-Maskawa (CKM) matrix. 
In Eq.~(\ref{BottomUp:2HDM:Yukawa2}), $\rho_f$ are the $3\times 3$ complex matrices which are not completely general unlike the general 2HDM due to the reality of the original Yukawa matrices. 
The matrices $\rho_f$ are constrained to be the following form: 
\begin{align}
\rho_f  &= \frac{\sqrt{2}}{s_{2\beta} v}\Bigg[\left(c_{2\beta} +\frac{i}{\tan(2I_f\xi)}\right) M_f \notag\\
&- \left(1 + \frac{i}{\tan(2I_f\xi)}\right)(V_f^\dagger V_f^* M_f U_f^TU_f)\Bigg], \label{eq:yf}
\end{align}
where $I_f = 1/2~(-1/2)$ for $f = u~(d,e)$, and 
$V_f$ and $U_f$ are the unitary matrices which appear via the bi-unitary transformation of fermions as $f_L \to V_f f_L$ and $f_R \to U_f f_R$ with the relation to the CKM matrix as $V= V_u^\dagger V_d$. 
The matrices $\rho_f$ given in Eq.~(\ref{eq:yf})
are generally not diagonal, so that they induce FCNCs via the neutral Higgs boson mediation at tree level. 
In order to avoid such FCNCs, the Yukawa alignment~\cite{Pich:2009sp} is often imposed, i.e., $Y_2^f = \tilde{\zeta}_f Y_1^f$ with $\tilde{\zeta}_f$ being an arbitrary real parameter. In this case, however, the mass matrices for fermions can be expressed by the real $3\times 3$ matrices with an overall complex phase factor in the SCPV 2HDM, so that the KM phase cannot be reproduced as the physical phase.   
Therefore, the {\it Yukawa misalignment} is necessary to reproduce the KM phase~\cite{Nierste:2019fbx}. 
This also means that a $Z_2$ symmetry cannot be imposed to the Higgs sector.

Although we cannot impose the Yukawa alignment on all the Yukawa interactions, we {\it can} impose it on the down-type quark and the charged lepton sectors. 
In this case, the mass matrices for down-type quarks and charged leptons are taken to be the real $3\times 3$ matrices with overall phases, so that  
the unitary matrices $V_{d,e}$ and $U_{d,e}$ can be rewritten as
\begin{align}
V_{d,e} = O_{d,e}^L,~~~~
U_{d,e} = O_{d,e}^R\,e^{-i\theta_{d,e}}, \label{eq:orthogonal}
\end{align}
where $O_{d,e}^{L,R}$ are the orthogonal matrices and
\begin{align}
 \theta_{d,e} =\arctan\left[\frac{t_\beta s_\xi \tilde{\zeta}_{d,e}}{1+t_\beta c_\xi \tilde{\zeta}_{d,e}}  \right].   
\end{align}
In this scenario, $\rho_f$ given in Eq.~(\ref{eq:yf}) take the following form:
\begin{align}
&\rho_u  = \frac{\sqrt{2}}{s_{2\beta} v}\left[\left(c_{2\beta} +\frac{i}{t_\xi}\right) M_u 
- \left(1 + \frac{i}{t_\xi}\right)(V V^T M_u U_u^TU_u)\right], \label{eq:rho_u}\\
&\rho_{d,e}  = \frac{\sqrt{2}M_{d,e}}{v}\zeta_{d,e}, 
\end{align}
where we used $V_u^\dagger V_u^*  = VV^T$, and 
\begin{align}
\zeta_{d,e} &= \frac{1}{s_{2\beta}}\left[\left(c_{2\beta} - \frac{i}{t_\xi}\right)
- \left(1 - \frac{i}{t_\xi}\right)e^{-2i\theta_{d,e}} \right] \notag\\
& = \frac{\tilde{\zeta}_{d,e}\,e^{i\xi}-t_\beta}{1 +  t_\beta \,\tilde{\zeta}_{d,e}\,e^{i\xi}}. 
\end{align} 
%

\noindent
{\it Non-decoupling extra Higgs bosons} -- As we have seen above, the masses of extra Higgs bosons purely come from the VEV.
This gives significant loop effects of the extra Higgs bosons on the observed Higgs boson couplings. 
We here especially consider those for the Higgs self-coupling $hhh$ and the loop-induced Higgs decays $h \to \gamma\gamma/Z\gamma$.  

The $hhh$ coupling is calculated by using the effective potential method at one-loop level, see Appendix~\ref{app:b} for details. 
In the Higgs alignment limit, the $hhh$ coupling subtracted by its SM prediction is expressed as
\begin{align}
\Delta \lambda_{hhh}^{\rm 1-loop} &\equiv (\lambda_{hhh}^{\rm 1-loop})_{\rm 2HDM} - (\lambda_{hhh}^{\rm 1-loop})_{\rm SM} \notag\\
&= 
 \frac{1}{32\pi^2}\sum_{\phi = H^\pm,H_2,H_3} n_\phi
\frac{\lambda^3_{\phi\phi h}}{m_\phi^2} \notag\\
& = \frac{1}{32\pi^2}\sum_{\phi = H^\pm,H_2,H_3} n_\phi\frac{(2m_\phi^2 + m_h^2)^3}{v^3m_\phi^2}, \label{eq:dellam}
\end{align}
where $n_{H^\pm}^{} =2~(1)$ for $\phi = H^\pm~(H_{2,3})$ and $\lambda_{abc}^{} \equiv \partial^3 V/(\partial a\partial b\partial c)|_0$ with $|_0$ representing all scalar fields being set to zero.
We see that the difference is enhanced by the quartic-power like dependence of the masses of extra Higgs bosons~\cite{Kanemura:2002vm,Kanemura:2004mg}.   

The decay rates of $h \to \gamma\gamma$ and $h \to Z\gamma$ are given as 
\begin{align}
&\Gamma_{h \to V\gamma}=  \frac{\sqrt{2}G_F\alpha_{\text{em}}^2m_h^3}{8\pi^3 (1+\delta_{V\gamma})} \left(1 - \frac{m_V^2}{m_h^2} \right)^3 \notag\\
& \times \left|(t,W)_V + \frac{v\lambda_{H^+H^-h}}{m_h^2 - m_V^2} c_V^{}F_V(m_{H^\pm})\right|^2, ~~
(V=Z,\gamma),
\end{align}
where 
$c_V^{} = 1~(\cot2\theta_W)$ with $\theta_W$ being the weak mixing angle for $V=\gamma~(Z)$. The $(t,W)_V$ term inside the squared absolute value denotes the top and W boson loop contributions whose numerical values are given to be about $1.6~(2.9)$ for $V=\gamma~(Z)$ in the SM case, see e.g. \cite{Gunion:1989we} for their analytic expressions. 
The loop function $F_V$ is given by  
\begin{align}
F_V(m)&=\frac{m_V^2}{2(m_h^2-m_V^2)}\Big[B_0(m_V^2;m,m) - B_0(m_h^2;m,m)\Big]\notag\\
&-\frac{1}{2}-m^2C_0(m_V^2,0,m_h^2;m,m,m),
\label{eq:loop-func}
\end{align}
with $B_0$ and $C_0$ being the Passarino-Veltman's 2-point and 3-point scalar functions~\cite{Passarino:1978jh}, respectively.
In the Higgs alignment limit, we have 
\begin{align}
   \lambda_{H^+H^-h} = \frac{1}{v}(2m_{H^\pm}^2 + m_h^2), \label{eq:lamhphmh} 
\end{align}
and the deviations of these decay rates purely come from 
the $H^\pm$ loop.  
For $m_h^2/m_{H^\pm}^2 \ll 1 $, 
the $H^\pm$ loop contribution asymptotically becomes $-1/12$ for $h \to \gamma\gamma$ and $-\cot2\theta_W/12$ for $h \to Z\gamma$, so that it gives a destructive interference with the $(t,W)_V$ part, and it makes the branching ratios of $h \to \gamma\gamma$ ($h \to Z\gamma$) to be about 10\% (4\%) smaller than the SM prediction.

\noindent
{\it Constraints} -- 
The scalar quartic couplings are constrained by the vacuum stability bound~\cite{Deshpande:1977rw,Nie:1998yn,Kanemura:1999xf}, the
tree-level perturbative unitarity bound~\cite{Kanemura:1993hm,Ginzburg:2005dt,Kanemura:2015ska}\footnote{The perturbative unitarity bound has been evaluated at next-leading order (NLO) in 2HDMs~\cite{Grinstein:2015rtl,Cacchio:2016qyh,Murphy:2017ojk}, and it has been applied to the SCPV 2HDM in Ref.~\cite{Nierste:2019fbx}. The upper limit on the masses of extra Higgs bosons has been given to be $m_{H^\pm}\lesssim 435$ (650) GeV, $m_{H_2}\lesssim 485$ (600) GeV and $m_{H_3}\lesssim 545$ (700) GeV by using the perturbative unitarity bound at NLO~\cite{Nierste:2019fbx} (leading order). } and the electroweak oblique $S$ and $T$ parameters~\cite{Peskin:1990zt,Peskin:1991sw}. In the general 2HDM, these $S$ and $T$ parameters have been calculated in~\cite{Grimus:2007if,Grimus:2008nb,Branco:2011iw,Asadi:2022xiy}. 
Due to the Yukawa misalignment,  $\rho_u$ given in Eq.~(\ref{eq:rho_u}) is strongly constrained by flavor observables, including meson mixings~\cite{UTfit:2023NP} and 
$b\to s\gamma$~\cite{Crivellin:2013wna,Altunkaynak:2015twa}. 
Direct searches at the LHC for charged~\cite{ATLAS:2021upq,CMS:2020imj} and neutral Higgs bosons~\cite{ATLAS:2023tlp,CMS:2023xpx,ATLAS:2024mih,CMS:2024ubt} provide additional bounds on $\rho_u$.
The CPV phase $\xi$ contributes to the electric dipole moment (EDM). We impose the upper bound on the electron EDM, $|d_e| \leq 4.1\times 10^{-30}$ e-cm (90\% CL)~\cite{Roussy:2022cmp}. 
For the constraints from the electron EDM, we compute the Barr-Zee diagrams~\cite{Barr:1990um} which give the dominant contribution to the EDM based on the formulae given in Ref.~\cite{Kanemura:2020ibp}. See also~\cite{Altmannshofer:2025nsl} for the recent calculation of the electron EDM including non Barr-Zee types. 

The $\rho_u$ matrix depends on the unitary matrix $U_u$ as shown in Eq.~(\ref{eq:rho_u}) which can be parameterized as follows:
\begin{align}
U_{u} &= R_{23}(\theta_{23},\delta_{23})R_{13}(\theta_{13},\delta_{13})R_{12}(\theta_{12},\delta_{12}), 
\end{align}
where $R_{ij}(\theta_{ij},\delta_{ij})$ are the SU(3) rotation matrices with $i'$--$i'$ ($i'\neq i,j$), $i$--$i$/$j$--$j$ and $i$--$j$ ($j$--$i$) components being 1, $\cos\theta_{ij}$ and $\sin\theta_{ij}e^{-i\delta_{ij}}$ ($-\sin\theta_{ij}e^{i\delta_{ij}}$), respectively, while all the other components being zero. 
In addition to these angles and phases of $U_u$, 
there are 8 free parameters in our model, i.e., 
$\{m_{H^\pm},m_{H_2},m_{H_3},\tan\beta,\tan\xi,\alpha,\tilde{\zeta}_{d,e}\}$. All these parameters and their ranges used for our scan are detailed in Tab.~\ref{tab:scan}. 
We randomly scanned 1 million parameter points, and found that approximately 1500 points satisfy all the bounds.

\begin{figure}[t]
 \begin{center}
\includegraphics[width=9.5cm]{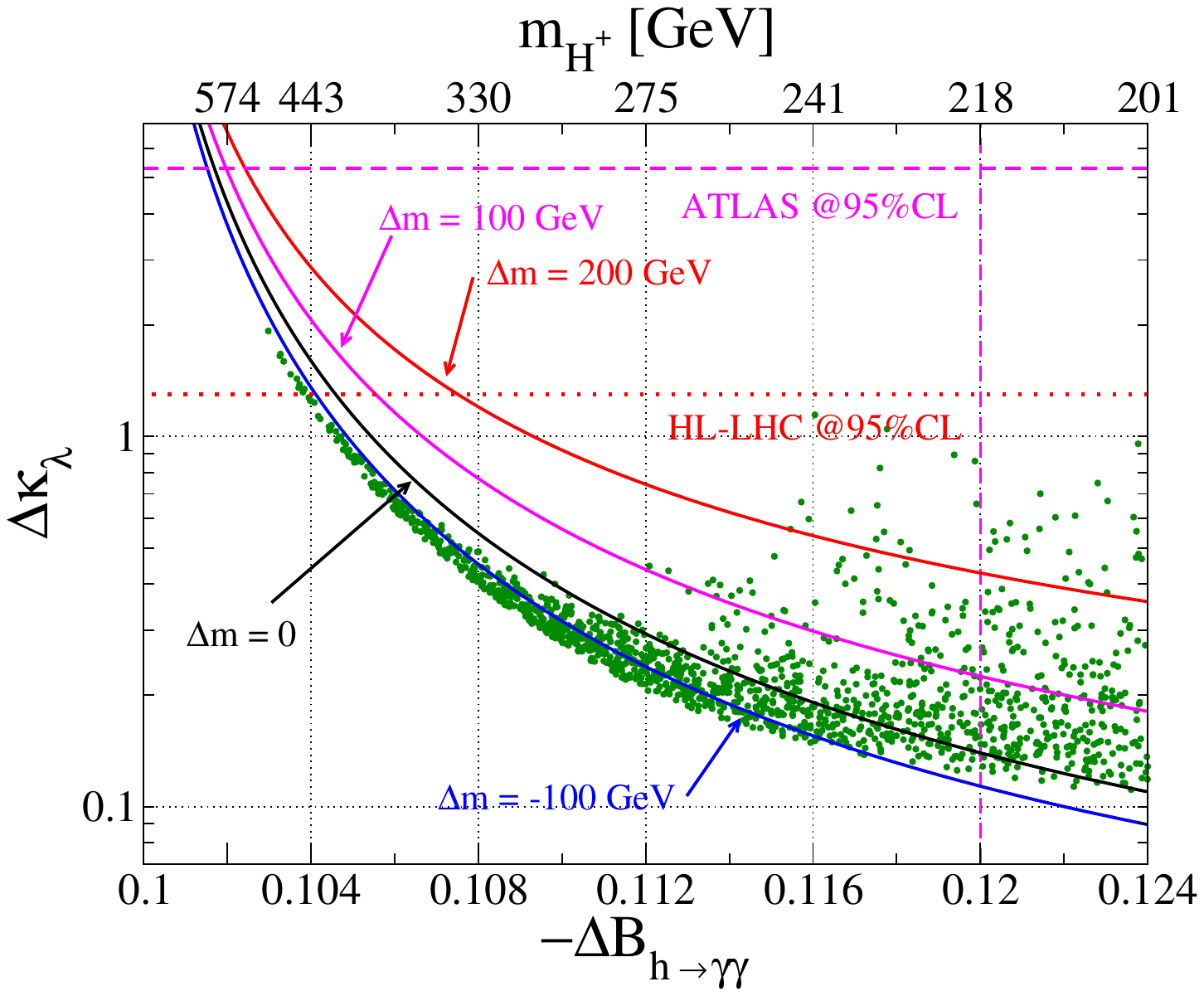} 
\caption{Correlation between $\Delta {\cal B}_{h\to \gamma\gamma}$ and $\Delta \kappa_\lambda$ defined in Eq.~(\ref{eq:delta}) in the Higgs alignment limit.  
The blue, black, magenta and red curves show the cases with $\Delta m (\equiv m_{H_2}^{} - m_{H^\pm}$)= $-100,$ 0, 100 and 200 GeV, respectively, and $m_{H^\pm} = m_{H_3}$ for all curves.    
The magenta-dashed lines represent the current upper limit   taken by the ATLAS experiments at 95\% CL, while the red-dotted line shows the expected upper limit on $\Delta \kappa_\lambda$ at 95\% CL given at the HL-LHC.  
The green dots show the predictions allowed by the theoretical constraints (perturbative unitarity and vacuum stability) and the experimental constraints ($S$, $T$ parameters, $b \to s\gamma$, meson mixings, electron EDM and direct searches for charged and extra neutral Higgs bosons at LHC). }
\label{fig:contourBR}
\end{center}
\end{figure}

\noindent
{\it Results} --
In Fig.~\ref{fig:contourBR}, 
we show our main result, the correlation between $\Delta {\cal B}_{h \to \gamma\gamma}$ and $\Delta\kappa_\lambda$, i.e., 
\begin{align}
\Delta {\cal B}_{h \to \gamma\gamma} &= 
\frac{{\cal B}_{h \to \gamma\gamma}^{\rm 2HDM}}{{\cal B}_{h \to \gamma\gamma}^{\rm SM}}-1, ~~
\Delta\kappa_\lambda  = \frac{\Delta\lambda_{hhh}^{\rm 1-loop}}{(\lambda_{hhh}^{\rm 1-loop})_{\rm SM}},\label{eq:delta}
\end{align}
with ${\cal B}_{h \to \gamma\gamma}^{\rm 2HDM(SM)}$ 
being the branching ratio of $h \to \gamma\gamma$ and 
$\Delta\lambda_{hhh}^{\rm 1-loop}$ given in Eq.~(\ref{eq:dellam}). 
%
The branching ratio of $h\to \gamma\gamma$ only depends on $m_{H^\pm}$ because of Eq.~(\ref{eq:lamhphmh}), so that we display the value of $m_{H^\pm}$ on the upper edge of the figure. From the ATLAS limit $-0.12 \leq \Delta {\cal B}_{h \to \gamma\gamma} \leq 0.18$~\cite{ATLAS:2022tnm}, we can extract $m_{H^\pm} \gtrsim 218$ GeV. 
The green points represent those satisfying the theoretical and experimental constraints discussed above.
We find that the lightest neutral extra Higgs boson, $H_2$, remains below the $t\bar{t}$ production threshold in all the benchmark points (BPs). 
For $m_{H^\pm}\gtrsim 300$ GeV, the masses of $H^\pm$ and $H_3$ tend to be nearly degenerate and larger than about 100 GeV than $m_{H_2}$ by the constraints. 
This leads to a quite strong correlation, e.g., ($\Delta {\cal B}_{h \to \gamma\gamma}$, $\Delta \kappa_\lambda$) = ($-10.3$\%,200\%) for $m_{H^\pm} \simeq$ 500 GeV and  ($-10.8$\%,50\%) for $m_{H^\pm} \simeq$ 330 GeV.   
The current constraint on $\Delta\kappa_\lambda$, i.e., 
$-1.4 \leq \Delta\kappa_\lambda \leq 5.3$~\cite{ATLAS:2022jtk,ATLAS:2022vkf}
is too weak to exclude the additional parameter space, but its future sensitivity at the HL-LHC is expected to be $\Delta\kappa_{\lambda} < 1.3$ at 95\% CL with 3 $\text{ab}^{-1}$ data~\cite{Cepeda:2019klc}. 
Thus, some of our BPs can be explored by the precise measurement of the $hhh$ coupling. 
On the other hand, $\Delta{\cal B}_{h \to \gamma\gamma}$ is expected to be measured with $4\%$ accuracy at 1$\sigma$ level~\cite{Cepeda:2019klc}, so that if the central value is given to be around 0, our scenario would be ruled out.

\noindent
{\it Phenomenology of extra Higgs bosons} -- In Tab.~\ref{tab:bp}, we present a few selected BPs that illustrate the characteristic
phenomenology in our model. 
As discussed above, the matrix $\rho_u$ is constrained from flavor physics, and $m_{H_2}$ remains below the $t\bar{t}$ production threshold. 
These constraints make the dominant decay mode of $H_2$ to be $H_2 \to tc$. 
Phenomenology of the heavier neutral Higgs boson ($H_3$) varies depending on the mass hierarchy of the extra Higgs bosons. In the degenerate case in masses, the decay of $H_3$ mimics that of $H_2$. However, in the non-degenerate case, $H_3$ decays into bosonic channels, $H_3 \to H_2 Z$ or $H^\pm W^\mp$, leading to different collider signatures. 
The decay of $H^\pm$ also strongly depends on $\rho_u$ and we find that along with the usual $tb$ mode,
$H^\pm$ decay substantially into unique albeit elusive $cb$ final state~\cite{Ghosh:2019exx,Hou:2024bzh}. In the case 
of non-degenerate spectrum, $H^\pm \to H_2 W^\pm$ also opens up which gives rise to distinctive 
phenomenology~\cite{Bahl:2021str,Mondal:2021bxa,Hou:2024ibt,Hou:2025tjp}.  

At the bottom of Tab.~\ref{tab:bp}, we also list the primary production channels for the extra Higgs bosons. The production 
cross-section for $H^\pm$, $\sigma(pp\to H^{\pm} tb)$, is computed at the leading order using 
\texttt{MadGraph5-aMC@NLO-3.5.3}\cite{Alwall:2011uj,Alwall:2014hca}. 
The heavy neutral Higgs bosons, $H_2$ and $H_3$, are produced predominantly via gluon fusion, and their cross-sections are estimated using \texttt{SusHi}~\cite{Harlander:2012pb,Harlander:2016hcx} 
at the NNLO level.

 \noindent
{\it Conclusion} -- Spontaneous CP violation provides a compelling framework for explaining the 
origin of CP violation beyond the Standard Model. The general two Higgs doublet model offers
 the simplest setting to realize this mechanism, leading to distinctive and testable predictions in 
 Higgs-sector observables. We have explored the non-decoupling property of the SCPV 2HDM,
  where the $hhh$ coupling and $h \to \gamma\gamma/Z\gamma$ decays significantly deviate 
  from the SM predictions. After taking into account all relevant theoretical and experimental 
  constraints, we have found that the branching ratio of $h \to \gamma\gamma$  ($h \to Z\gamma$) decreases more than 10\% (4\%) for 
  most of the parameter space, and the current signal strength for the $h \to \gamma\gamma$ mode gives a robust restriction on the mass of charged Higgs boson to be larger than 220 GeV.
    Such a large deviation in $h \to \gamma\gamma$ will be explored at the HL-LHC in the near future.

   In this Letter, we have demonstrated for the first time the complementarity of the $hhh$ coupling and 
   the $h\to\gamma\gamma$ decay in testing the scenarios with SCPV. 
   This interplay provides a 
   novel, robust strategy to explore the heavy scalar spectrum and our analysis sets new 
   exclusion bounds on heavy Higgs spectrum that up to now were in agreement with all 
   relevant constraints. We also demonstrated the characteristic decay patterns of the heavy 
   Higgs bosons such as $H_2 \to tc$ and $H^\pm \to cb$, which can serve as additional 
    experimental targets. We therefore strongly advocate systematic investigations of these
     non-standard channels in current and upcoming experiments such as the HL-LHC, as 
     they offer crucial opportunities to test the SCPV paradigm.

 \noindent
{\it Acknowledgments} -- T.M. was supported by BITS Pilani Grant NFSG/PIL/2023/P3801. 

\bibliography{references}

\newpage

\begin{appendix}

\section{Higgs boson masses \label{app:a}} 

The mass matrix for the neutral Higgs bosons ${\cal M}$ (symmetric $3\times 3$ form) can be written in the Higgs basis, i.e., ($h_1',h_2',h_3'$) as
\begin{align}
\begin{split}   
\frac{{\cal M}_{11}}{v^2} & = \lambda_1 c_\beta^4 + \lambda_2s_\beta^4
+2(\lambda_3 +\lambda_4 + \lambda_5 c_{2\xi}  )c_\beta^2 s_\beta^2 \\
&+2(\lambda_6c_\beta^2 + \lambda_7 s_\beta^2)s_{2\beta}c_\xi, \\
\frac{{\cal M}_{22}}{v^2} & = (\lambda_1 + \lambda_2 -2 \lambda_3  -2 \lambda_4 )c_\beta^2s_\beta^2  + \frac{\lambda_5}{2}(1 + c_{2\beta}^2c_{2\xi}) \\
&- (\lambda_6 - \lambda_7)s_{2\beta}c_{2\beta}c_\xi, \\
\frac{{\cal M}_{33}}{v^2} & = \lambda_5 s_\xi^2, \\
\frac{{\cal M}_{12}}{v^2} & = (\lambda_2s_\beta^2 - \lambda_1c_\beta^2)c_\beta s_\beta + \frac{1}{4}(\lambda_3 + \lambda_4 + \lambda_5 c_{2\xi})s_{4\beta}\\
& + [\lambda_6(2c_{2\beta} - 1)c_\beta^2 + \lambda_7(2c_{2\beta} + 1)s_\beta^2] c_\xi, \\
\frac{{\cal M}_{13}}{v^2} & = -\frac{\lambda_5}{2}s_{2\beta}s_{2\xi}-(\lambda_6c_\beta^2 + \lambda_7s_\beta^2)s_\xi, \\
\frac{{\cal M}_{23}}{v^2} & =-\frac{1}{2}\lambda_5 c_{2\beta}s_{2\xi} + \frac{1}{2}(\lambda_6 - \lambda_7)s_{2\beta}s_\xi. 
\end{split}
\label{eq:masses}
\end{align}
The Higgs alignment condition, ${\cal M}_{12} = {\cal M}_{13}=0$, can then be solved in terms of the $\lambda_{6,7}$ as follows: 
\begin{align}
\begin{split}
\lambda_6 &= \Big[-\lambda_1c_{\beta} +\lambda_2t_\beta s_\beta + (\lambda_3 + \lambda_4)\frac{c_{2\beta}}{c_\beta} \\
&+\frac{\lambda_5}{c_\beta}(1 - 2c_{\beta}^2(2 + c_{2\xi}))\Big]\frac{s_\beta}{2c_\xi}, \\
\lambda_7 &= \Big[ \lambda_1 \frac{c_\beta}{t_\beta} -\lambda_2 s_\beta -(\lambda_3 + \lambda_4) \frac{c_{2\beta}}{s_\beta} \\
&+ \frac{\lambda_5}{s_\beta}(1- 2s^2_\beta(2+c_{2\xi}))\Big]\frac{c_\beta}{2c_\xi}. 
\end{split} \label{eq:ha}
\end{align}

\section{Effective potential \label{app:b}} 

The effective potential is given by \begin{align}
V_{\rm eff} = V_0 + V_1 + V_{\rm CT}, 
\end{align}
where $V_0$, $V_1$ and $V_{\rm CT}$ are respectively the contributions from the tree level diagrams, the one-loop diagrams and the counterterms. 
Assuming no charge-breaking vacuum, we can parameterize the background fields in the Higgs basis as 
\begin{align}
\Phi = \begin{pmatrix}
0 \\
\frac{\varphi_1}{\sqrt{2}}
\end{pmatrix},~~
\Phi' = \begin{pmatrix}
0 \\
\frac{\varphi_2 + i\varphi_3}{\sqrt{2}}
\end{pmatrix}.  \label{eq:bg}
\end{align}
We note that $\varphi_{2,3}$ are generally nonzero even in the Higgs basis due to the radiative corrections. 
We then obtain $V_0$ by using Eqs.~(\ref{eq:hb2}), (\ref{pot_thdm1})  and (\ref{eq:bg}). 
The counterterm $V_{\rm CT}$ can be obtained by replacing $m_i^2 \to \delta m_i^2$ and $\lambda_j \to \delta \lambda_j$ in $V_0$.

The one-loop contribution is given by~\cite{Coleman:1973jx,Dolan:1973qd}
\begin{align}
V_1
&= \sum_{\phi}V_1^{\phi} + \cdots ~~(\phi = H^\pm,~H_{1,2,3})\notag\\
&= 
\frac{n_\phi}{64\pi^2}m_{\phi}^4(\varphi)
 \left[
 \ln \left(\frac{m_{\phi}^2(\varphi)}{\mu^2} \right) -c \right] 
 + \cdots,  \label{eq:V1}
\end{align}
where $m_{\phi}^2(\varphi)$ denote the field dependent masses of $\phi$, $n_\phi = 2~(1)$
for $\phi = H^\pm~(H_{1,2,3})$, and the ellipsis show the other contributions from SM fermion and gauge boson loops. 
In the above expression, $\mu^2$ is an arbitrary dimensionful parameter, and   
$c$ is a constant including ultra-violet divergence given by 
\begin{align}
c = \frac{3}{2} + \frac{1}{\epsilon} -\gamma_E^{} + \ln 4\pi, 
\end{align}
with $\epsilon$ being defined via the $D$ dimensional spacetime $D = 4 - 2\epsilon$ in the dimensional regularization and $\gamma_E^{}$ being the Eular-Mascheroni constant.   

We impose the renormalization conditions~\cite{Enomoto:2021dkl}
\begin{align}
    (V_1 + V_{\rm CT})_{\varphi_i}
    = 0, \quad 
     (V_1 + V_{\rm CT})_{\varphi_i\varphi_j}
    = 0, 
\end{align}
where we introduced the following shorthand notation of the derivatives: 
\begin{align}
X_{\varphi_i} \equiv \frac{\partial X}{\partial \varphi_i}\Bigg|_{v},\quad  
X_{\varphi_i\varphi_j} \equiv \frac{\partial^2 X}{\partial \varphi_i\partial \varphi_j}\Bigg|_v, 
\end{align}
with $|_v$ denoting the replacement of $\{\varphi_1,\varphi_2,\varphi_3\}\to \{v,0,0\}$ after the derivative.
These conditions determine 9 counterterms, and ensure that the vacuum and masses are not changed from those given at tree level. 
In particular, if we take the third derivative with respect to $\varphi_1$, we obtain 
\begin{align}
&V_{{\rm eff},\,\varphi_1\varphi_1\varphi_1}  = 
\lambda_{h_1'h_1'h_1'}
+ \frac{1}{32\pi^2}\sum_\phi n_\phi
\frac{[(m_\phi^2)_{\varphi_1}]^3}{m_\phi^2} 
\notag\\
&+\sum_\phi\left[\frac{3}{v^2} + \frac{(m_\phi^2)_{\varphi_1\varphi_1\varphi_1}}{(m_\phi^2)_{\varphi_1}} 
-\frac{3[(m_{\phi}^2)_{\varphi_1\varphi_1}]^2}{[(m_\phi^2)_{\varphi_1}]^2}\right]V_{1,\,\varphi_1}^\phi \notag\\
&+ \sum_\phi\left[-\frac{3}{v} +\frac{3(m_\phi^2)_{\varphi_1\varphi_1}}{(m_\phi^2)_{\varphi_1}} \right]V_{1,\,\varphi_1\varphi_1}^\phi + \cdots
\label{eq:vppp}.
\end{align}
We note that the first derivative of the field dependent mass provides the corresponding scalar trilinear coupling, i.e., $(m_\phi^2)_{\varphi_i} =\lambda_{\phi\phi^* \varphi_i}$. 

In the Higgs alignment case, 
we have the following relations 
\begin{align}
(m_\phi^2)_{\varphi_1 \varphi_i} &= \lambda_{\phi\phi^* h_1'h_i'} = \frac{\lambda_{\phi\phi^* h_i'}}{v}, ~
(m_\phi^2)_{\varphi_1 \varphi_i\varphi_j} =0.
\end{align}
Therefore, the second and third lines of Eq.~(\ref{eq:vppp}) vanish, and we obtain the simple form given in Eq.~(\ref{eq:dellam}).  

\section{Benchmark points}

\begin{table}[b]
\begin{center}
\begin{tabular}{|c|c||c|c|}
\hline
Parameters  & Range   & Parameters    & Range  \\ \hline\hline
$m_{H^{\pm}/H_{2}/H_{3}}$(GeV) & 200 -- 800 & $\alpha$         & $-\pi/2$ -- $\pi/2$ \\ \hline
$\tan\beta$         & 1 -- 8     & $\theta_{12/23/13}$& 
$0$ -- $\pi$   \\ \hline
$\tan\xi$       & 0.5 -- 50  & $\delta_{12/23/13}$ & $0$ -- $2\pi$  \\ \hline
$\tilde{\zeta}_d$       & 0 -- 50  & $\tilde{\zeta}_e$ &  1 \\ \hline
\end{tabular}
\caption{Parameters and their scan ranges used for this analysis.}
\label{tab:scan}
\end{center}
\end{table}

We show the list of the parameters scanned in the numerical analysis and their ranges in Tab.~\ref{tab:scan}. 
The BPs discussed in the main text are shown in Tab.~\ref{tab:bp}.

\begin{table*}[!t]
\begin{tabular}{|c|c|c|c|c|}
\hline
Parameters                          & BP1                  & BP2                  & BP3                      & BP4                  \\ \hline
$m_{H^{\pm}/H_{2}/H_{3}}$(GeV)      & 455 / 228 / 454       & 264 / 212 / 223       & 236 / 248 / 446           & 246 / 243 / 245       \\ \hline
$\tan\beta$                         & 2.61                  & 3.12                  & 2.51                      & 2.53                  \\ \hline
$\tan\xi $                          & 33.13                 & 7.95                  & 13.45                     & 42.31                 \\ \hline
$\alpha ~(\text{rad})$                     & 0.0                   & $-$0.5                  & $-$0.05                     & 0.35                  \\ \hline
$\tilde{\zeta}_d/\tilde{\zeta}_e$               & 32/1                  & 38/1                  & 41/1                      & 31/1                  \\ \hline
$\theta_{12/23/13} ~(\text{rad})$          & 1.64 / 0.43 / 1.55   & 1.71 / 1.67 / 1.51 & 1.55 / 3.13 / 1.49       & 1.56 / 2.72 / 1.45   \\ \hline
$\delta_{12/23/13}~(\text{rad})$           & 1.93 / 6.02 / 1.60    & 5.41 / 0.88 / 1.69    & 2.02 / 4.47 / 1.64        & 1.11 /  3.04 / 1.60   \\ \hline
\multirow{3}{*}{${\cal B}_{H_{2}\to XY}$ (\%) }   & $tc:82.93$            &$tc:96.85$             & $ tc:99.04$               & $tc:99.56$            \\ 
                                    & $\tau\tau:4.70$       & $\tau\tau:1.03$       & $\tau\tau:0.23$           & --                    \\ 
                                    & $ tu:3.88$            & $tu:0.62$             & $ tu:0.21$                & --                    \\ \hline
\multirow{3}{*}{${\cal B}_{H_3\to XY}$ (\%)}    & $H_{2}Z:89.52$        & $tc:97.34$            & $H^{\pm}W^{\mp}:66.46$    & $tc:99.5$             \\ 
                                    & $tt:9.57$             & $\tau\tau:0.73$       & $H_{2}Z:25.98$            & --                    \\ 
                                    & $tc:0.59$             & $tu:0.63$             & $tt:3.78$                 & --                    \\ \hline
\multirow{3}{*}{${\cal B}_{H^\pm\to XY}$ (\%)}  & $H_{2}W^{\pm}:78.1$   & $tb:56.78$            & $cb:61.28$                & $cb:77.95$            \\ 
                                    & $tb:21.23$            & $cb:43.11$            & $tb:38.67 $               & $tb:22.03$            \\ 
                                    & $cb:0.65$             &$ H_{2}W^{\pm}:0.06$   & $\tau\nu:0.03$            & --                    \\ \hline
$\sigma(pp\to H^{\pm} tb/H^{\pm} cb)~(\text{pb})$       & 0.087/0.088                  & 0.253/6.367                 & 0.471/19.762                    & 0.371/38.671                 \\ \hline
$\sigma(pp\to H_{2/3})~(\text{pb})$          & 4.277 / 4.247         & 4.854 / 7.022         & 3.702 / 4.657             & 4.163 / 8.348         \\ \hline
\end{tabular}
\caption{BPs satisfying all theoretical and experimental constraints.}
\label{tab:bp}
\end{table*}

\end{appendix}

\end{document}